\documentclass[12pt]{article}
\usepackage{graphicx}
\title{\bf THE COLLIDING WINDS OF WR146: SEEING THE WORKS}
\author{
E.~P.~O'Connor $^{1,2}$, S.~M.~Dougherty $^2$, J.~M.~Pittard$^3$, 
P.~M.~Williams$^4$\\
\vspace{1cm}\\
\normalsize $^1$Physics Dept., U. Prince Edward Island, Charlottetown, PEI., Canada \\
\normalsize $^2$National Resarch Council, Herzberg Institute, DRAO., Penticton, Canada \\
\normalsize $^3$Dept. Physics and Astonomy, U. Leeds, Leeds, UK.\\
\normalsize $^4$Institute for Astronomy, U. Edinburgh, Scotland, UK.\\
}
\date{\mbox{}}

\newcommand\aj{{AJ}}%
\newcommand\apj{{ApJ}}
\newcommand\apjs{{ApJS}}%
\newcommand\aap{{A\&A}}%
\newcommand\mnras{{MNRAS}}%

\begin{document}
\maketitle
\pagestyle{empty}
%
%
\def\bull{\vrule height .9ex width .8ex depth -.1ex}
\makeatletter
\def\ps@plain{\let\@mkboth\gobbletwo
\def\@oddhead{}\def\@oddfoot{\hfil\tiny\bull\quad
``Massive Stars and High-Energy Emission in OB Associations''; JENAM
2005, held in Li\`ege\ (Belgium)\quad\bull}%
\def\@evenhead{}\let\@evenfoot\@oddfoot}
\makeatother
%
%
\def\beginrefer{\section*{References}%
\begin{quotation}\mbox{}\par}
\def\refer#1\par{{\setlength{\parindent}{-\leftmargin}\indent#1\par}}
\def\endrefer{\end{quotation}}
%
%
{\noindent\small{\bf Abstract:} WR146 is a WC6+O8 colliding-wind
binary (CWB) system with thermal emission from the stellar winds of
the two stars, and bright non-thermal emission from the wind-collision
region (WCR) where the winds collide. We present high resolution radio
observations from 1.4 to 43 GHz that give one of the best quality
radio spectra of any CWB to date.  Observations at 22 GHz now span 8
years, and reveal the proper motion of the system, allowing comparison
of multi-epoch data.  VLBI observations show the location of the WCR
relative to the stellar components, from which the wind momentum ratio
can be shown to be $\eta=0.06\pm0.15$. The radio spectrum and the
spatial distribution of emission are modelled, and we determine the
contribution of both stellar winds and the WCR to the observed
emission. We show that our current models fail to account for the high
frequency spectrum of WR146, and also produce too much emission far
from the stagnation point of the wind collision. }
%
%
\section{Observations}
On 2004 October 1, WR\thinspace146 was observed at 1.4, 4.8, 8.3, 15,
and 43 GHz with the highest-resolution configuration of the Very Large
Array (VLA) in conjunction with the Pie Town antenna of the Very Large
Baseline Array (VLBA). Two components are observed at 15, 22, and 43
GHz.  At 43~GHz, the two sources are separated by $152\pm2$~ mas
(Fig. 1 - left). The northern component (N) has a predominately
synchrotron spectrum at almost all frequencies, except at 43 GHz where
it is a combination of emission from the WCR and the O star stellar
wind (see Fig. 3). The southern source (S) has a thermal spectrum and
we identify this as the wind from the WC6 star (Dougherty et al. 1996,
2000), which STIS spectroscopy confirms is the spectral type of the
southern component.

WR146 was also observed at 4.9 GHz using the European VLBI Network
(EVN) and the MERLIN array in the UK on 2001 February 12.  The EVN
observation shows a bow-shaped arc of emission (Fig.~1 - centre), with
a brightness temperature of $10^7$~K, which we identify as the
WCR. Such a shape is consistent with the WCR ``wrapping'' around the
star with the lower wind momentum, the northern O star (e.g. see
Eichler \& Usov 1993, Dougherty et al. 2003). The EVN observations
indicate the WR and O star are south and north of the WCR, consistent
with previous suggestions (Dougherty et al. 1996, 2000; L{\'e}pine et
al. 2001). MERLIN also detects only the WCR.
\begin{figure}
  \includegraphics[width=5.5cm]{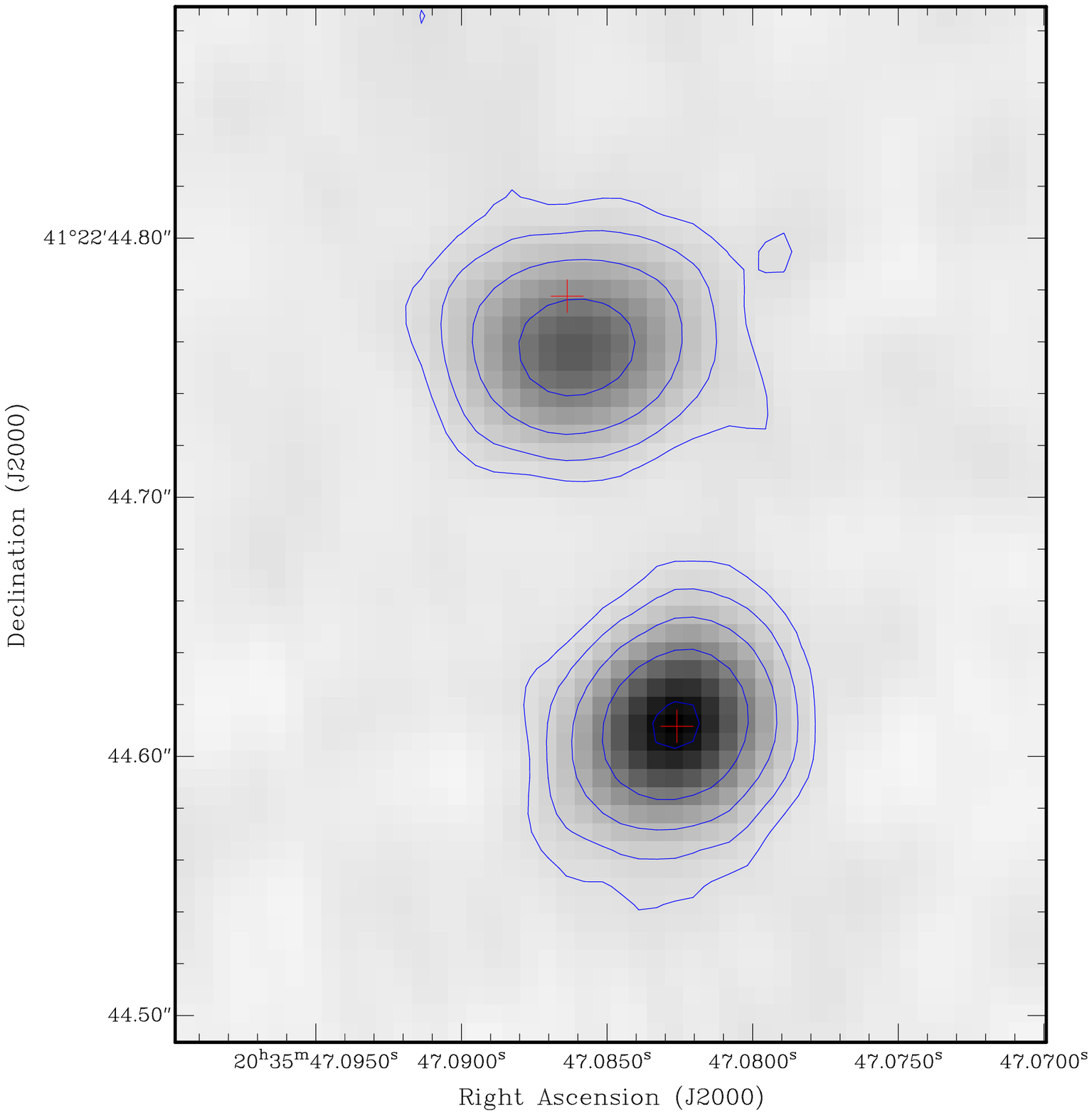}
  \includegraphics[width=5.5cm]{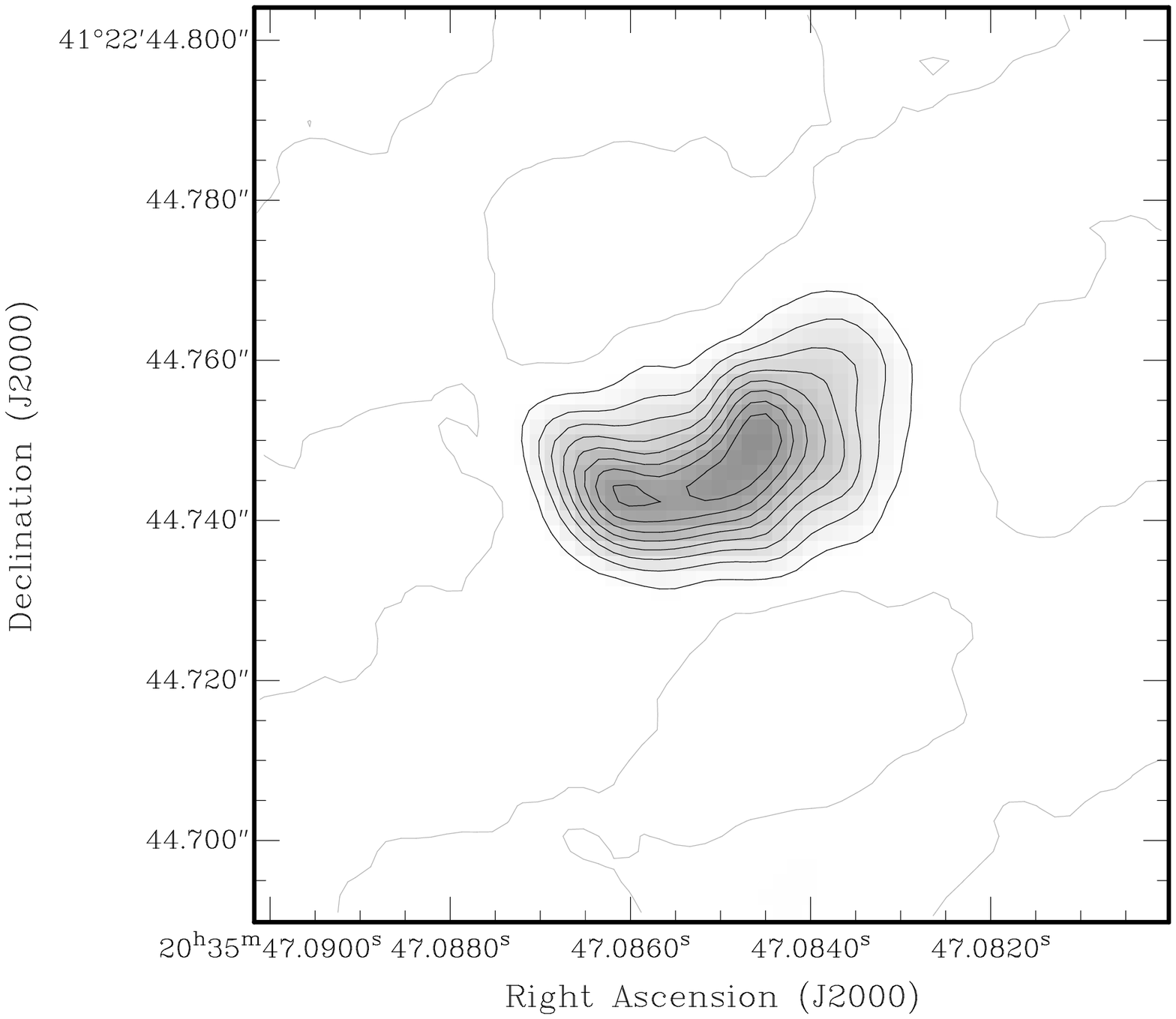}
  \includegraphics[width=5.5cm]{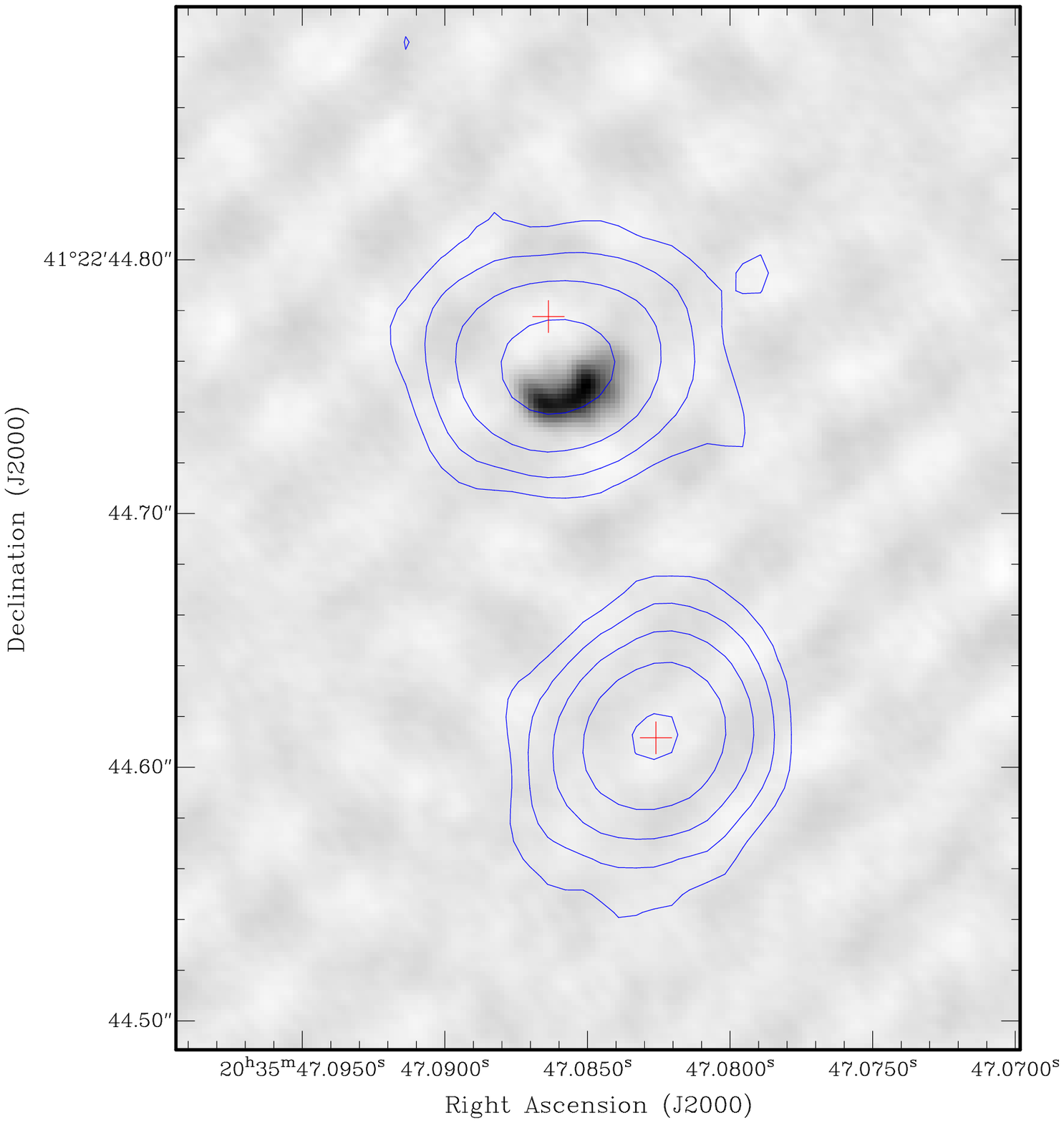}
  \caption{(Left) WR146 with VLA+PT at 43 GHz, with resolution of 30
mas.  Two sources are clearly observed, with a separation of $152\pm2$
mas. (Centre) EVN observation at 5 GHz with resolution of 9~mas. A
bow-shaped region is observed as expected for a WCR. (Right) An
overlay of the VLA 43 GHz (contours) with the EVN (greyscale) 5-GHz
emission shown. The proper motion of the source between the
observations has been taken into account. The stellar positions are
marked, as deduced from HST observations, with a separation of
168~mas.}
\end{figure}

\section{Proper Motion}
In order to align correctly radio images of any stellar source
obtained at widely separated epochs it is necessary to know the proper
motion of the source. In addition to the VLA observations from 2004,
WR\thinspace146 was also observed at 22 GHz with the VLA on 1996
October 26 and 1999 August 26.  Each of these observations was
phase-referenced using the same quasar, J2007+404, and the change in
the relative position of WR\thinspace146 and J2007+404 can be
determined over the 8 years of observations.  A weighted regression
fit to the relative positions of WR146 leads to a proper motion of
$\mu_\alpha=-3.65\pm0.17$ and $\mu_\delta=-6.46\pm 0.40$
mas~yr$^{-1}$.
\begin{figure}
  \centering
  \includegraphics[width=8.3cm]{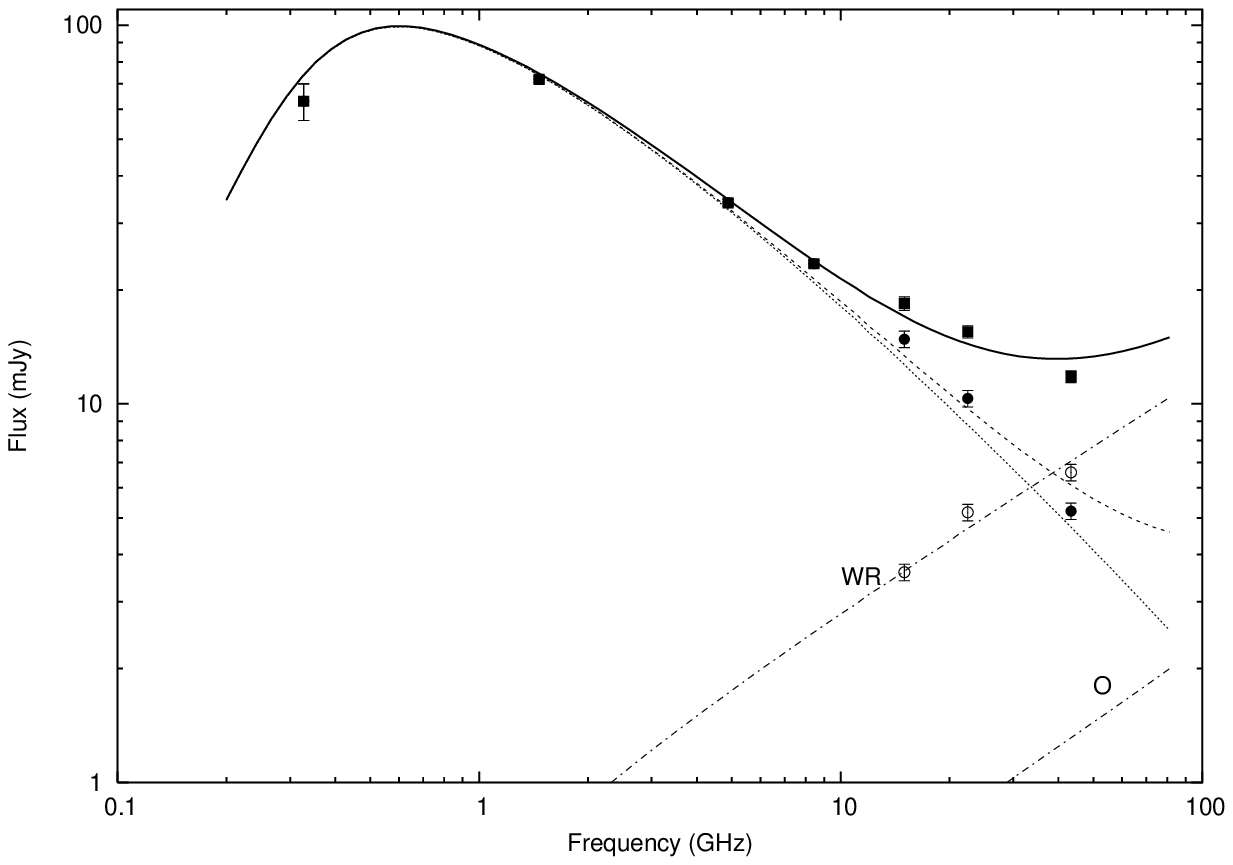}
  \includegraphics[width=8.3cm]{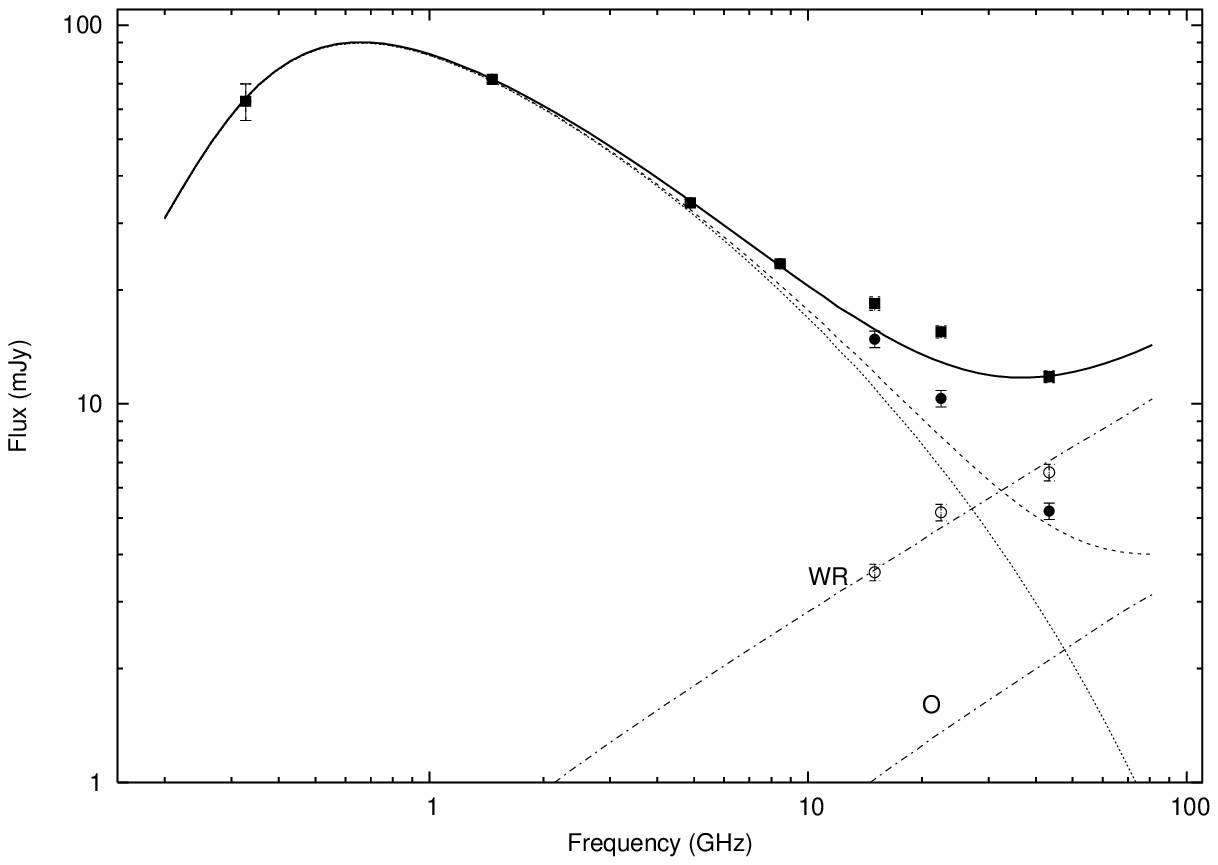}
  \caption{Comparison of observations and two model spectra of
WR146. Solid squares mark the total flux from N+S (327~MHz datum from
Taylor et al. (1996)), solid circles the flux from N and open circles
that from S.  Best-fit models of the emission from component N
(synchrotron - dotted; thermal - lower dot-dashed; synchrotron+thermal
- dashed) and the thermal emission from S (higher dot-dashed) are
shown. The total flux from N+S is shown by the solid line. The two
models presented are where the high frequency synchrotron spectrum is
determined by IC cooling (left) or the high energy cutoff of the shock
accelerated electrons (right).}
\end{figure}

\section{Component separation and wind-momentum ratio}
By observing the relative position of the stars and the WCR (located
around the stagnation point of the two stellar winds), we can
determine the wind-momentum ratio $\eta$.  After taking account of the
proper motion, a comparison of the EVN and VLA observations reveal
the position of the WCR relative to the WR star, with a separation
of $135\pm6$~mas. From Niemela et al. (1998), the HST observed a
separation of $168\pm31$~mas between the WR and O stars. Taking the
uncertainty of the location of the stagnation point as the half-width
of the WCR as detected by the EVN, we find $\eta=0.06\pm0.15$. The
bulk of the error is due to the uncertainty of the stellar separation
($\pm31$~mas). 

\section{Modelling the Spectrum}
We have modelled the radio emission from WR\thinspace146 using a
hydro-dynamical model of the density and pressure distribution of a
colliding-wind system and solving the radiative transfer equation for
thermal and magneto-bremsstrahlung emission and absorption, as well
accounting for the Razin effect, inverse-Compton (IC) and Coulomb
cooling (Dougherty et al. 2003, Pittard et al. 2005).  The model
successfully recovers the total flux up to 8.3 GHz.  At higher
frequencies the fit is less convincing, due to a poor model of
component N. Either the 43-GHz flux is overestimated or the 15 and
22-GHz fluxes are underestimated (see Fig. 2). These problems can be
resolved if the synchrotron spectrum steepens around $\sim 22$~GHz. 

\section{Simulated observations}
Using the model parameters derived from the spectrum we can generate
synthetic images of WR146, using the AIPS routine UVCON.  This allows
us to ``observe'' our model (Fig.~3), and impose constraints on the
model from the spatial distribution of the model emission.  The
similarities between the synthetic images and the observations is
remarkable, particularly at 43~GHz. Clearly, the 43-GHz emission is a
combination of O-star stellar wind and WCR emission - its peak is not
located at either the contact discontinuity or the stellar position,
but between them. Comparison with Fig. 1 (left) suggests the emission
from N is too extended, both E-W and N-S. The EVN model image also
reveals the spatial distribution of the model emission is too wide.
We conclude that the model produces too much emission far from the
stagnation point of the WCR. Our models also show that deducing the
location of the stagnation point from observations is complicated by
the relative brightness of the WR and O-star shocked plasma in the
WCR.

\begin{figure}
  \centering
  \includegraphics[width=7cm]{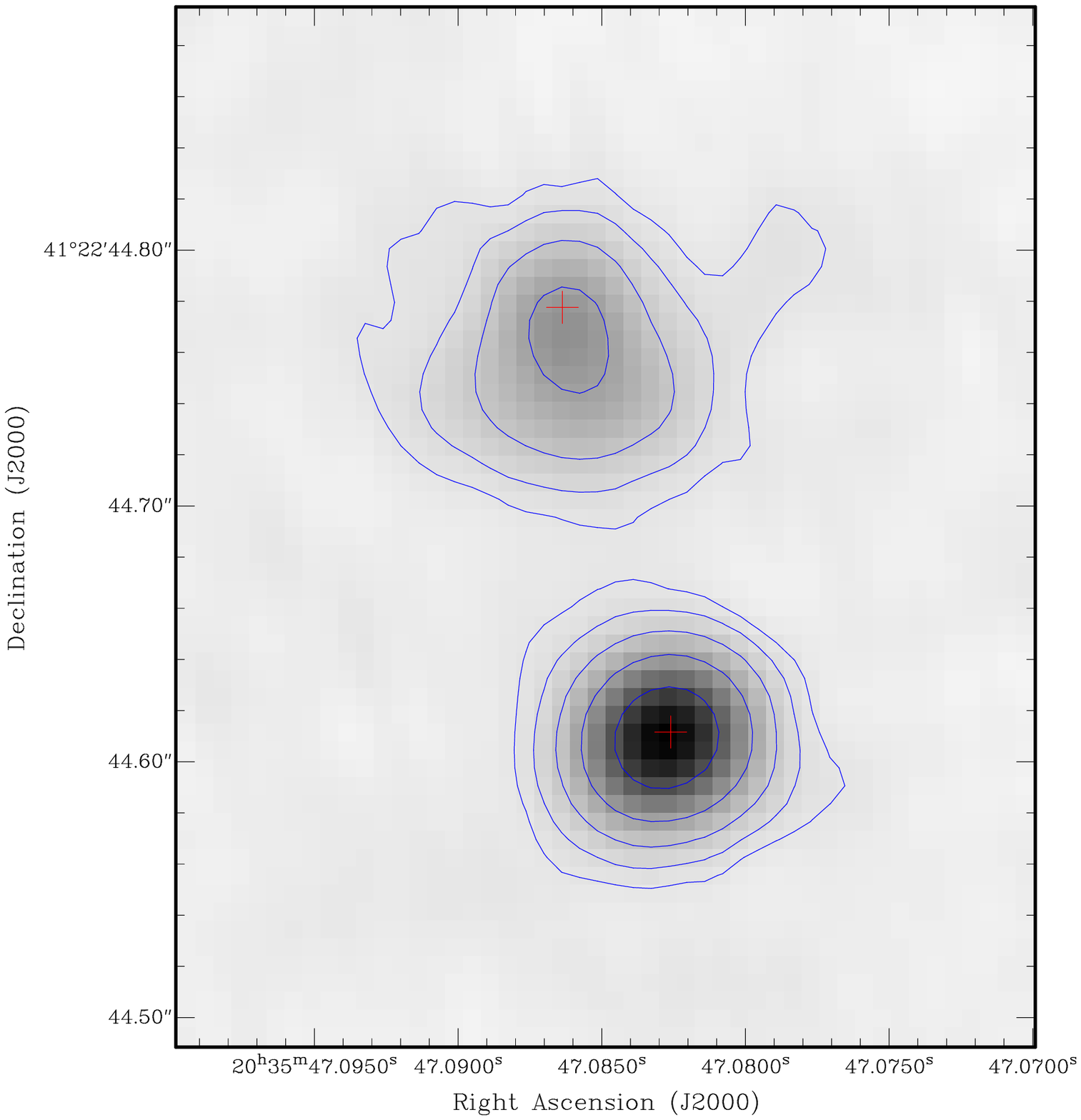}
  \includegraphics[width=7cm]{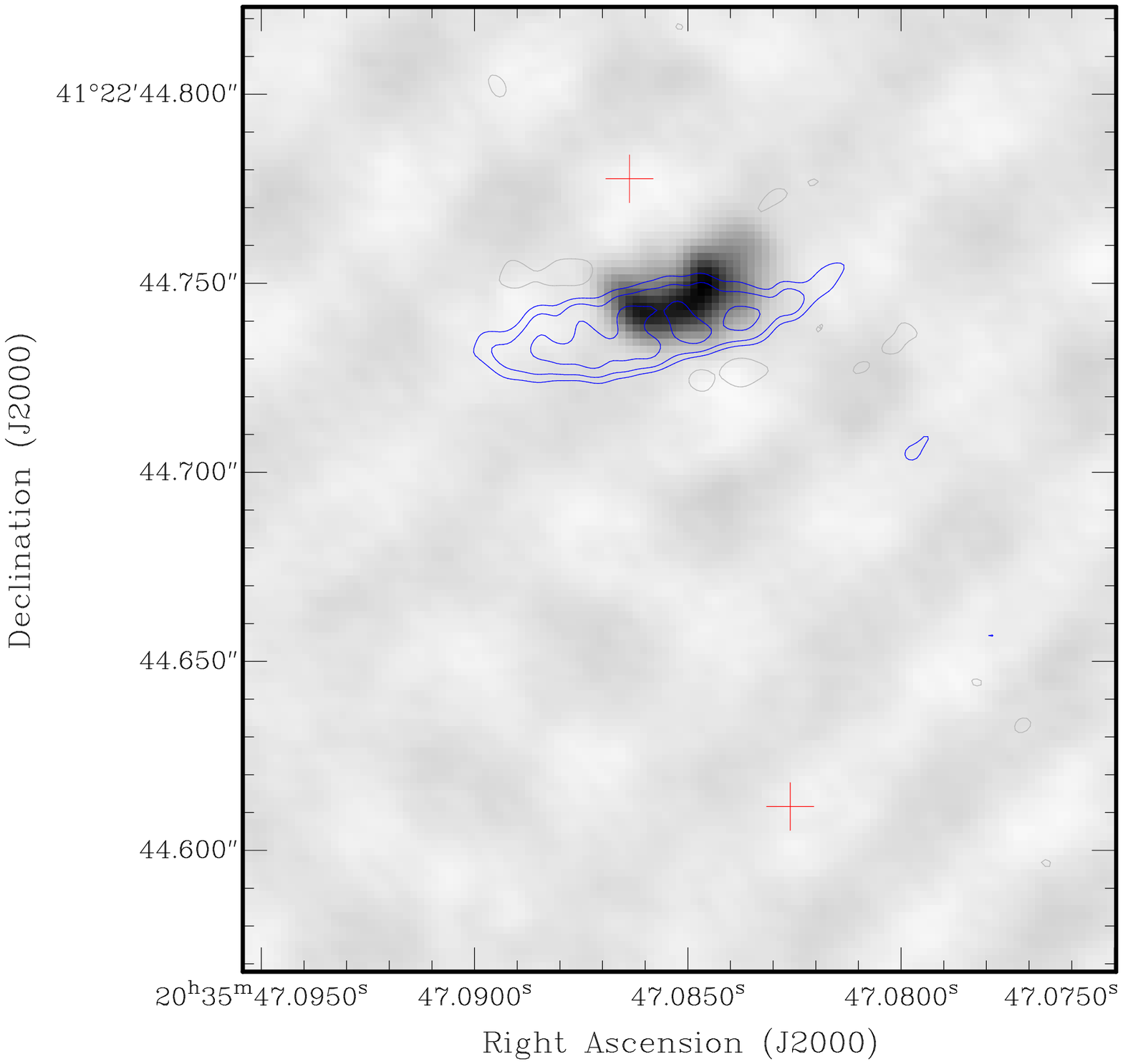}
  \caption{Simulated observations of WR146 at 43 GHz with the VLA+PT
(left), and at 4.9 GHz with the EVN (right). The input 43-GHz model is
based on the model from Fig. 2 (right), and the EVN model from Fig. 2
(left). The crosses mark the relative location of the two stars as
determined by HST observations .  Compare the left-hand figure to
Fig. 1 (left). The EVN observation and simulated data are superimposed
for a clear comparison.}
\end{figure}

%
%
\section*{Acknowledgments}
This work was supported by the National Research Council of
Canada, and the University of Prince Edward Island Co-op programme.
JMP acknowledges gratefully funding from the Royal Society. The
observations were obtained from the Very Large Array,
operated by the National Radio Astronomy Observatory, and the
MERLIN array, operated by the University of Manchester on behalf of
PPARC. The European VLBI Network is a joint facility of European,
Chinese, South African and other radio astronomy institutes funded by
their national research councils.

%
%
 
\beginrefer
\refer Dougherty, S.~M., Beasley, A.~J., Claussen, M.~J., \& Zauderer, B.~A.,
  Bolingbroke,~N.~J. 2005, \apj, 623, 447

\refer Dougherty, S.~M., Pittard, J.~M., Kasian, L., Coker, R.~F., Williams,
  P.~M., \& Lloyd, H.~M. 2003, \aap, 409, 217

\refer Dougherty, S.~M., Williams, P.~M., \& Pollacco, D.~L. 2000, \mnras, 316,
  143

\refer Dougherty, S.~M., Williams, P.~M., van der Hucht, K.~A., Bode, M.~F.,
  \& Davis, R.~J. 1996, \mnras, 280, 963

\refer Eichler, D. \& Usov, V. 1993, \apj, 402, 271  

\refer L\' epine, S.~., Wallace, D., Shara, M.~M., Moffat, A.~F.~J., \&
  Niemela, V.~S. 2001, \aj, 122, 3407

\refer Niemela, V.~S., Shara, M.~M., Wallace, D.~J., Zurek, D.~R., \&
  Moffat, A.~F.~J. 1998, \aj, 115, 2047

\refer Pittard, J.~M., Dougherty, S.~M., Coker, R., O'Connor, E.,
  \& Bolingbroke, N.~J. 2005, \aap, submitted

\refer Taylor, A.~R., Goss, W.~M., Coleman, P.~H., van Leeuwen, J., \&
  Wallace, B.~J. 1996, \apjs, 107, 239

\endrefer           

\end{document}